\documentstyle[epsfig]{aipproc}

\begin{document}
\title{Simulations of black hole--neutron star binary coalescence}

\author{William H. Lee} \address{Instituto de Astronom\'{\i}a, UNAM \\
Apdo. Postal 70--264, Cd. Universitaria M\'{e}xico D.F. 04510 MEXICO \\
wlee@astroscu.unam.mx}

\lefthead{Lee}
\righthead{BH-NS coalescence}
\maketitle

\begin{abstract}

We show the results of dynamical simulations of the coalescence of
black hole--neutron star binaries. We use a Newtonian Smooth Particle
Hydrodynamics code, and include the effects of gravitational radiation
back reaction with the quadrupole approximation for point masses, and
compute the gravitational radiation waveforms. We assume a polytropic
equation of state determines the structure of the neutron star in
equilibrium, and use an ideal gas law to follow the dynamical
evolution. Three main parameters are explored: (i) The distribution of
angular momentum in the system in the initial configuration, namely
tidally locked systems vs. irrotational binaries; (ii) The stiffness
of the equation of state through the value of the adiabatic index
$\Gamma$ (ranging from $\Gamma=5/3$ to $\Gamma=3$); (iii) The initial
mass ratio $q=M_{NS}/M_{BH}$. We find that it is the value of $\Gamma$
that determines how the coalescence takes place, with immediate and
complete tidal disruption for $\Gamma \leq 2$, while the core of the
neutron star survives and stays in orbit around the black hole for
$\Gamma=3$. This result is largely independent of the initial mass
ratio and spin configuration, and is reflected directly in the
gravitational radiation signal. For a wide range of mass ratios,
massive accretion disks are formed ($M_{disk}\approx 0.2 M_{\odot}$),
with baryon--free regions that could possibly give rise to gamma ray
bursts.

\end{abstract}

\section{Introduction}

The emission of gravitational waves in a binary system will eventually
drive the system to coalesce through angular momentum losses, if the
decay time is less than the Hubble time. The observed decay for known
binary neutron star systems such as PSR~1913+16 matches the prediction
of general relativity to high accuracy (Stairs et al.~1998). There are
still no observations of black hole--neutron star systems, but they
are believed to exist, with corresponding coalescence rates that are
comparable to those of double NS systems (see Kalogera \&
Belczy\'{n}ski, these proceedings).

These systems are candidates for detection by gravitational wave
detectors such as LIGO and VIRGO. Although the final coalescence
signal will probably be out of the frequency range of the first
observatories, the inspiral phase, during which the stars can be
thought of as point masses, will certainly be observable in this
respect. For the final coalescence waveform, modeling the
hydrodynamics in the system becomes an important issue, as it can
affect its evolution in a significant manner. For example, Newtonian
tidal effects due to the finite size of the stars, can alone
de--stabilize the orbit and make it decay on a dynamical timescale
(see Lai, Rasio \& Shapiro~1993a).

We have studied the dynamical interactions in close black
hole--neutron star binary systems previously for a variety of initial
configurations (Lee \& Klu\'{z}niak~1995, 1999a,b, hereafter papers~I
\&~II respectively; Lee~2000, hereafter paper~III), and present here
an overview of the results these simulations have produced. The
results are not only relevant for the production of gravitational
waves, but also for the progenitor systems of gamma ray bursts (GRBs,
Klu\'{z}niak \& Lee~1998; Janka et al.~1999), and the production of
heavy elements through r--process nucleosynthesis (Lattimer \&
Schramm~1974, 1976; Symbalisty \& Schramm~1982).

\section{Numerical method and initial conditions}

\subsection{Numerical method}

All the computations presented here have been carried out using the
Smooth Particle Hydrodynamics (SPH) method. This is a Lagrangian
technique, originally developed by Lucy~(1977) and Gingold \&
Monaghan~(1977) and is ideally suited for the study of complicated
flows in three dimensions, when no assumptions about symmetry in the
system are made, and where there are large volumes that are basically
devoid of matter. An excellent review has been given by
Monaghan~(1992). The code is essentially Newtonian, and makes use of a
binary tree structure to find hydrodynamical forces and carry out the
gravitational force computation, with a multipole expansion to
quadrupole order. The viscosity is artificial, its purpose being the
modeling of shocks and avoiding the interpenetration of SPH
particles. We have settled on the form of Balsara~(1995) for our
latest work, since it minimizes the effects of shear viscosity on the
evolution of the system. This is of particular importance for this
work, since massive accretion disks are often formed around the black
hole after the initial dynamical encounter.

The black hole is modeled as a Newtonian point mass, producing a potential 
\begin{equation}
\Phi=-GM_{BH}/|\vec{r}-\vec{r}_{BH}|
\end{equation}
at position $\vec{r}$. To model the horizon, an absorbing boundary is
placed at the Schwarzschild radius $r_{Sch}=2GM_{BH}/c^{2}$. Any
particle crossing this boundary is removed from the simulation. Its
mass is added to that of the black hole, and the latter's position and
momentum are adjusted so as to ensure the conservation of total mass
and total linear momentum in the system. The simulations shown here
have used between 8,000 and 40,000 SPH particles initially. Since
accretion onto the black hole entails a loss of particles, this number
decreases during the simulation.

In most of the simulations presented here we have included a back
reaction term to mock the effect that the emission of gravitational
waves has on the evolution of the system, by draining angular momentum
during the orbital evolution. This acceleration is calculated in the
quadrupole approximation, assuming the two components are point masses
(see e.g. Zhuge, Centrella \& McMillan~1996; Davies et al.~1994;
Rosswog et al.~1999). This term in the equations of motion is switched
off when (and if) complete tidal disruption occurs, or if the mass of
the secondary (neutron star) core drops below a certain limit (usually
one tenth of the initial neutron star mass), in the cases when the
star is not completely shredded by tidal forces (details of the
implementation of the back reaction force can be found in Paper~III).

The gravitational wave emission is calculated in the quadrupole
approximation, by adding the contribution from the fluid as a whole to
that of the black hole as a point mass. This gives the waveforms
directly from the second derivatives of the inertia tensor. To obtain
the luminosity an additional (numerical) derivative is required. For
the calculation of power spectra, one can attach a point--mass
inspiral signal to the coalescence waveform at earlier times, that
matches smoothly at the time the dynamical simulation is started.

\subsection{Initial Conditions}\label{initial}

To perform dynamical simulations, we first construct a neutron star in
hydrostatic equilibrium. The equation of state is taken to be that of
a polytrope, with $P=K\rho^{\Gamma}$ ($K$ and $\Gamma$ are taken
constant throughout the star). We place $N$ SPH particles on a cubic
lattice, with masses proportional to the Lane--Emden density at the
corresponding radius. This ensures that the spatial resolution is
approximately constant, and it helps to model the edge of the star
more accurately, since that is where the density gradient is often the
largest, for the values of the adiabatic index $\Gamma $ that we
consider. In all the calculations presented here, the neutron star has
mass $M_{NS}=1.4 M_{\odot}$ and radius $R_{NS}=13.4$~km. For each
value of the index $\Gamma=3; 2.5; 2; 5/3$ we find the value of $K$
that ensures this mass and radius. After placing the particles on the
cubic lattice, the star is relaxed in an inertial reference frame,
with an artificial damping term in the equations of motion, keeping
the specific entropy constant. After relaxation, all our spherical
stars satisfy the virial theorem to within one part in $10^{3}$. For
the following, distances and masses are measured in units of $R_{NS}$
and $M_{NS}$, so that time and density are measured in units of
$t=1.146\times10^{-4}s(R/R_{NS})^{3/2}(M_{NS}/1.4M_{\odot})^{-1/2}$
and
$\rho=1.14\times10^{18}kg~m^{-3}(R/R_{NS})^{-3}(M_{NS}/1.4M_{\odot})$

The choice of a polytropic equation of state (instead of a physical
equation of state such as that of Lattimer \& Swesty~1991) was made in
order to explore what effect the compressibility of the fluid has on
the global evolution of the system, and on the gravitational wave
signal. For polytropes, the mass--radius relationship is $R \propto
M^{(\Gamma-2)/(3\Gamma-4)}$. So this means that for $\Gamma> 2$, the
neutron star will respond to mass loss by shrinking, while for $\Gamma
<2$ it will expand. A star with $\Gamma=2$ has a radius that is
independent of its mass. As we will show below, this has a crucial
effect on the outcome of the coalescence process.

The next step in constructing the initial conditions depends on the
spin of the star. Typically, the binary separation is only a few
stelar radii when we begin our dynamical calculations, and so the
tidal deformation of the neutron star is quite large. We consider two
extreme cases of angular distribution in the system. 

In the first, the star is tidally locked, so that the same side of the
star always faces the black hole companion. This initial condition is
easy to set up, since the system is in a state of rigid rotation. Thus
we can view the binary in the co--rotating frame and neglect Coriolis
forces (since we are interested in an equilibrium configuration), with
an artifical damping term in the equations of motion and wait for it
to relax to a static configuration. While this relaxation procedure is
carried out, the orbital velocity of the co--rotating frame is
continously adjusted so that the force on the center of mass of the
star is exactly balanced by the centrifugal acceleration.

The second type of initial condition we consider is that of an
irrotational binary. In this case, the star has essentially zero spin
when viewed from an external, inertial reference frame. This is more
complicated to set up, and we have used the method of Lai, Rasio \&
Shapiro~(1993b) for our calculations. They developed a variational
method to obtain a solution to this problem by approximating the star
as a tri--axial ellipsoid (in this case an irrotational Roche--Riemann
ellipsoid). It still experiences tidal deformations, but the shape of
the star is fixed in the co--rotating frame, while internal motions
with zero circulation take place in its interior.

Realistically, the first kind of initial condition was shown to be
nearly impossible by Kochanek~(1992) and Bildsten \& Cutler~(1992),
because the viscosity inside neutron stars is not large enough to
maintain synchronization during the inspiral phase. Essentially, the
stars will coalesce with whatever spin configurations they have when
inspiral begins.

\begin{figure} 
\centerline{\epsfig{file=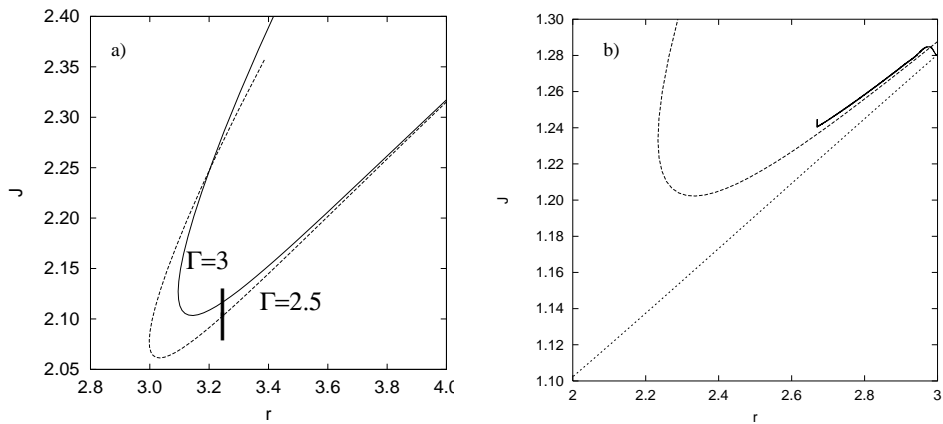,height=2.5in,width=5.0in}}
\vspace{6pt}
\caption{Total angular momentum $J$ as a function of binary separation
$r$ for various black hole--neutron star binaries. (a) Irrotational
binaries with $q=0.5$ and $\Gamma=3$ (solid line) and $\Gamma=2.5$
(dashed line). The thick black vertical line marks the separation used
to start dynamical calculations. (b) Tidally locked binary with
$\Gamma=5/3$ and $q=1$. The solid line is the result of an SPH
relaxation sequence, the dashed line results from approximating the
neutron star as a compressible tri--axial ellipsoid, and the dotted
line from assuming it is a rigid sphere. }
\label{fig1}
\end{figure}

In either case, the fact that the stars are {\em not} point masses has
a direct impact on the evolution of the system, and on its
configuration immediatly before coalescence. We show in Figure~1 plots
of total angular momentum $J$ in the system as a function of orbital
separation for several systems. It is clear that there are important
deviations from Keplerian point--mass behavior, and even from the
result obtained by treating the stars as rigid spheres. The turning
points in the curves show the presence of a dynamical instability,
which, once reached, can drive orbital decay on a dynamical timescale
(for many of the parameters in the runs shown here, this can be as
fast as the decay due to the emission of gravitational waves). It is
crucial to model the hydrodynamics in these systems if one is to
extract information from the gravitational wave signal produced during
coalescence.

We show in Table~1 the initial parameters for several of the dynamical
runs we have performed. We include the value of the adiabatic index,
the spin configuration, the mass ratio, the initial separation and the
number of particles initially used to model the neutron star.

\begin{table}
\caption{Initial set of parameters for dynamical runs.}
\label{table1}
\begin{tabular}{llccccc}
Run &   $\Gamma$ & Spin\tablenote{L: tidally locked; I: irrotational} & $q=M_{NS}/M_{BH}$ & $r_{i}/R_{NS}$ & N & Reference \\
\tableline
AL1.0 & 3.0 & L &  1.00 & 2.78 & 16,944 & Paper~I \\
AL0.31 & 3.0 & L &  0.31 & 3.76 &  8,121 & Paper~I \\
AI0.5 & 3.0 & I &  0.50 & 3.25 & 38,352 & Paper~III \\
AI0.31 & 3.0 & I &  0.31 & 3.76 & 38,352 & Paper~III \\
BI0.31 & 2.5 & I &  0.31 & 3.70 & 37,752 & Paper~III \\
CI0.31 & 2.0 & I & 0.31 & 3.70 & 37,560 & In preparation \\
DL1.0 & 5/3 & L &  1.00 & 2.70 & 17,256 & Paper~II \\
DI0.31 & 5/3 & I & 0.31 & 3.60 & 38,736 & In preparation \\
\end{tabular}
\end{table}

\section{Results}

\subsection{Systems with a stiff equation of state}

For systems with stiff equations of state ($\Gamma=3$ and
$\Gamma=2.5$) the star will respond to mass loss by shrinking
slightly, as mentioned above, and tidal effects are more pronounced
than for soft equations of state. This is because the star is not as
centrally condensed, and the moment of inertia is relatively large (a
substantial amount of the star's mass can be found in its outer
layers). This effect alone can be large enough to make the orbits
dynamically unstable for large enough mass ratios ($q\geq 0.5$, see Lai
Rasio \& Shapiro~1993a; Paper~I). 

In any event, angular momenum losses to gravitational waves make the
separation decrease, and Roche Lobe overflow occurs. Generally, the
orbital evolution of the system is similar for the tidally locked and
irrotational cases. A mass transfer stream forms from the neutron star
core to the black hole, and there is a rapid episode of accretion,
lasting a few milliseconds (peak accretion rates reach a few solar
masses per millisecond). What happens next depends on the value of the
adiabatic index.

For $\Gamma=3$, the core of the neutron star responds to mass loss by
shrinking enough to cut off the mass transfer stream, and it always
survives as a distinct body, remaining in an elliptical orbit around
the black hole, with a mass ranging from 0.2 to 0.3 solar masses (see
Figure~2). For initial mass ratios $q \geq 0.5$ a massive accretion disk
forms, containing a few tenths of a solar mass orbiting at a typical
distance of 100~km. For lower mass ratios, there is essentially no
disk at our level of resolution (only a handful of SPH particles,
amounting to $10^{-3}$ solar masses are in orbit around the black
hole). The orbital eccentricity is $e \simeq 0.2$, and the orbital
separation at each subsequent periastron passage is sufficiently small
so as to allow secondary episodes of mass transfer to occur, with the
gas being either directly accreted by the black hole in the absence of
a disk, or feeding it if one was formed during the initial encounter.

\begin{figure}[t] 
\centerline{\epsfig{file=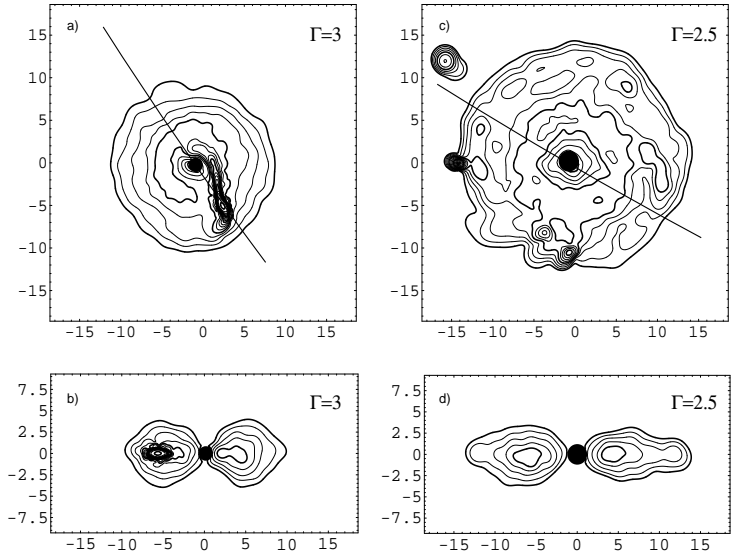,height=3.4in,width=3.8in}}
\vspace{6pt}
\caption{Density contours at the end of the dynamical simulations for
runs AI0.5 and BI0.31 in (a,c): the orbital plane, and (b,d): in the
meridional planes shown in (a,c) by the black lines. All contours are
logarithmic and equally spaced every 0.25 dex. Bold contours are
plotted at $\log \rho=-5,-4,-3,-2,-1$ (if present), in the units
defined in section~\ref{initial}.}
\label{fig2}
\end{figure}

For $\Gamma=2.5$, the neutron star is almost completely disrupted
during the initial encounter, with a small core surviving until the
second periastron passage a few milliseconds later. Tidal disruption
is then complete, and there is always a thick accretion torus around
the black hole containing approximately 0.2--0.3 solar masses. This
figure is fairly independent of the initial mass ratio (see Figures~2
and 3a).

During the encounters in irrotational binaries, long tidal tails of
material are formed (see Figure~3b). This gas is violently ejected
through the outer Lagrange point, and some of it (on the order of
$10^{-2}$--$10^{-1}$ solar masses) has enough mechanical energy to
escape the black hole + accretion torus system. This may be relevant
for the production of heavy elements through r--process
nucleosynthesis (see Rosswog et al.~1999; Freiburghaus, Rosswog \&
Thielemann~1999). It is interesting to note that these tidal tails are
practically nonexistent for the case of the tidally locked binaries
(see Figure~5b). This is because the latter events are less violent
during the initial encounter and mass transfer episode.
\begin{figure} 
\centerline{\epsfig{file=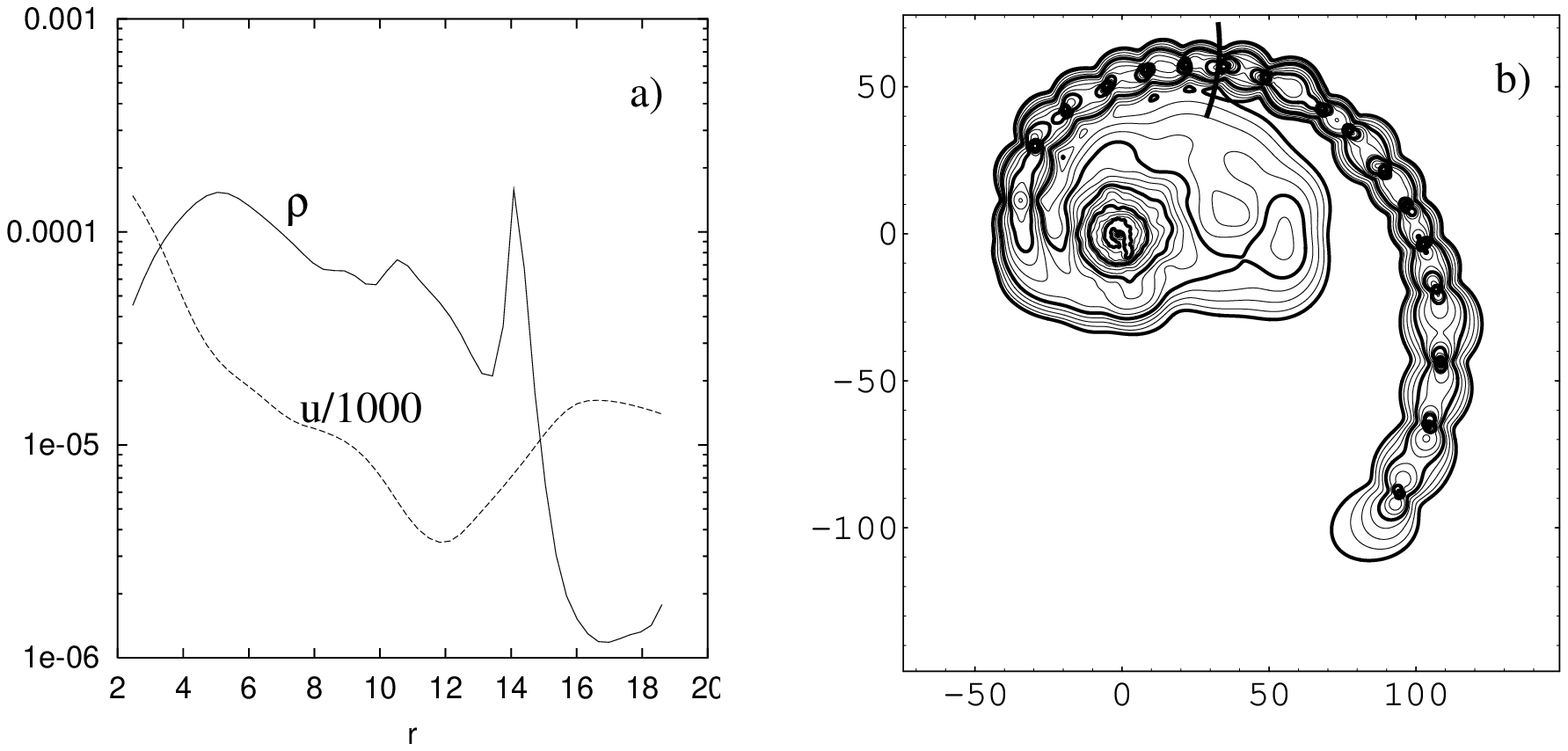,height=2.5in,width=5.5in}}
\vspace{6pt}
\caption{(a) Azimuthally averaged profiles for run BI0.31 in the
equatorial plane for the density $\rho$ and the specific internal
energy $u$ ($u/1000$ is plotted). (b) Density contours in the orbital
plane at the end of run AI0.5. All contours are logarithmic and
equally spaced every 0.25 dex. Bold contours are plotted at $\log
\rho=-8,-7,-6,-5,-4$, in the units defined in
section~\ref{initial}. The thick black line across the tidal tail
divides gas that is bound to the black hole from that which is on
outbound trajectories. }
\label{fig3}
\end{figure}
This can also be seen by comparing the orbital separation
during the coalescence and the secondary episodes of mass
transfer. For example, the final separation is on the order of
$7~R_{NS}$ for run AI0.31 (irrotational, see Paper~III), but only
$4.7~R_{NS}$ for run AL0.31 (tidally locked, see Paper~I). The only
difference between these two runs is the inital spin configuration,
they both had an initial mass ratio $q=0.31$ and initial separation
$r=3.76R_{NS}$. The amount of mass transferred from the neutron star
core to the black hole during each successive periastron passage is
lower in the irrotational case by approximately one order of
magnitude. The irrotational encounters are more violent because once
the separation becomes small enough, the tidal bulge on the neutron
star becomes larger and in a sense, tries to spin up the star. This
angular momentum can only come from the orbital component, and thus
the decay is faster.

We note that although the star is not immediately disrupted and
subsequent mass transfer events occur, this is not a stable or steady
state process at all. Gravitational radiation emission tends to
decrease the separation, and mass transfer tends to increase it (since
the donor is the less massive of the two components and the transfer
itself is almost conservative). But these processes appear to balance
each other through distinct events and not in a continous fashion (the
impossibility of stable mass transfer in such a system was pointed out
by Bildsten \& Cutler~1992 and Kochanek~1992).

The evolution described above determines what the gravitational
radiation signal is like. Since a system with azimuthal symmetry will
not radiate gravitational waves, any disk structure will not
contribute to such a signal. The one--armed spirals formed through the
ejection of gas from the system do not contain enough mass to
contribute significantly either, and so the signal at late times is
determined by the fate of the neutron star core (see Figure~4). If
there is complete tidal disruption, the signal essentially vanishes,
whereas if the binary survives, a persistent waveform will remain,
albeit with an lower amplitude and frequency (since the binary
separation has increased and the mass ratio has dropped as well).
\begin{figure}[t] 
\centerline{\epsfig{file=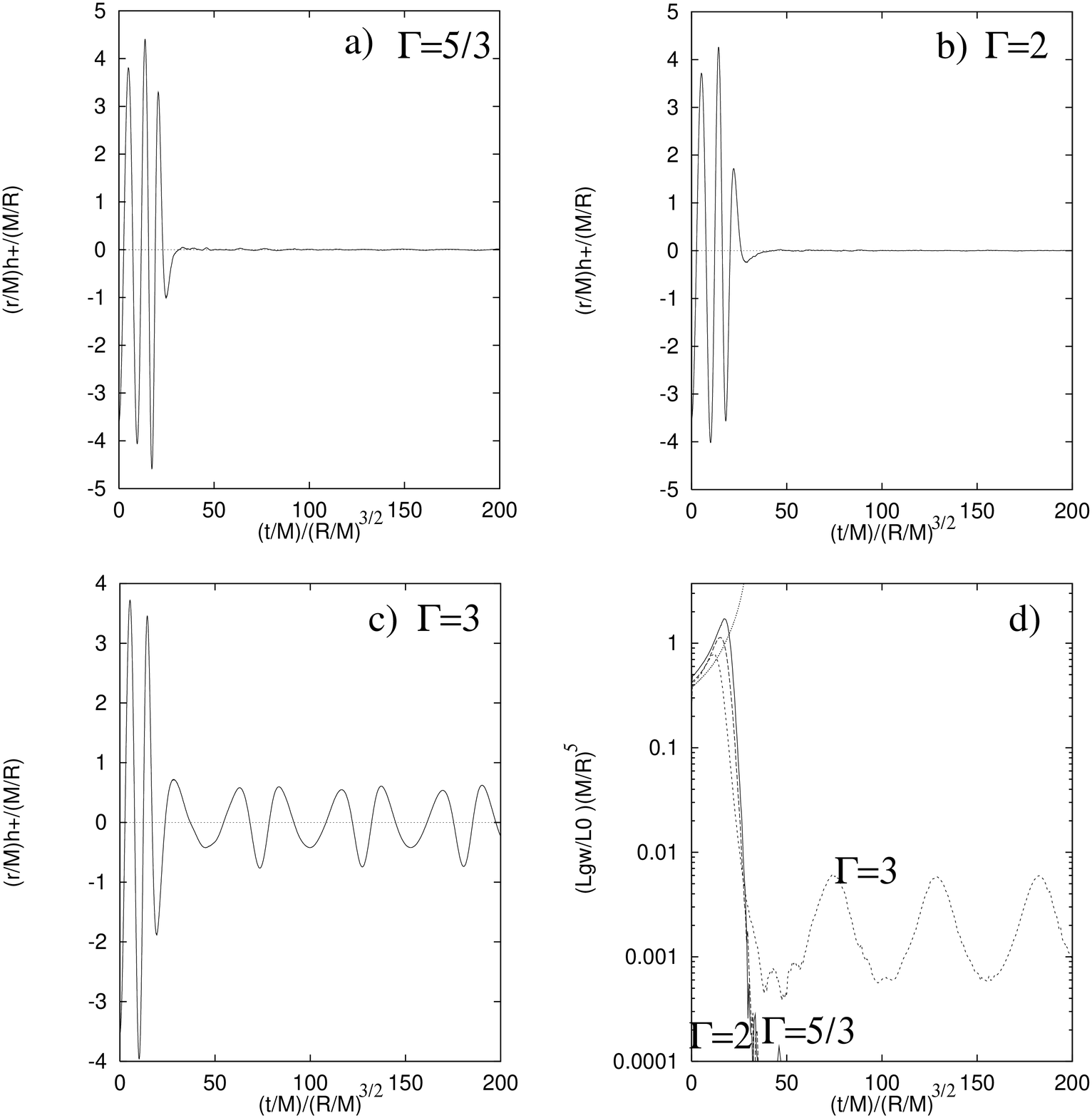,height=3.5in,width=4.5in}}
\vspace{6pt}
\caption{Gravitational radiation waveforms (one polarization) for an
observer located at a distance $r$ from the system, along the z--axis
(perpendicular to the orbital plane) for (a)~$\Gamma=5/3$ (run
DI0.31), (b)~$\Gamma=2$ (run CI0.31), (c)~$\Gamma=3$ (run AI0.31). In
(d) the luminosity of gravitational radiation is plotted for the same
cases. The monotonically increasing curve is the result for two point
masses, computed in the quadrupole approximation. }
\label{fig4}
\end{figure}
This is the case
for $\Gamma=3$ for a wide range of mass ratios. The frequency at which
the signal amplitude drops marks the onset of intense mass transfer
and is a function of the radius of the neutron star. In our Newtonian
calculations, this is outside of the LIGO band (between 800~Hz and
1200~Hz). However, general relativistic effects will likely make the
orbit unstable at larger separations, and thus the drop would happen
at lower frequencies, possibly within the range of LIGO (see Faber \&
Rasio for binary neutron star calculations using a post Newtonian
treatment on this point, these proceedings).

\subsection{Systems with a soft equation of state}

As before, for systems with a soft equation of state, angular momentum
losses to gravitational waves make the binary separation decrease
until the star overflows its Roche lobe. This in turn leads to
complete tidal disruption, regardless of the initial mass ratio (see
Figure~5a). A massive accretion disk is formed, with a few tenths of a
solar mass orbiting the black hole at a typical distance of 100~km.  A
large one--armed spiral forms, with some matter being ejected from the
system, both for the irrotational and tidally locked cases (the amount
is similar to that observed for the stiff equations of state, between
$10^{-2}$ and $10^{-1}$ solar masses). It is the soft equation of
state, with the ensuing mass--radius relationship, that causes this
behavior, as opposed to that observed for larger values of
$\Gamma$. For $\Gamma=5/3$, our softest index, mass loss by the
neutron star makes it expand, overflowing its Roche lobe even
further. The mass transfer process itself is unstable, and it can be
strong enough to de--stabilize the orbit by itself, excluding the
effects of gravitational radiation back reaction.

\begin{figure} 
\centerline{\epsfig{file=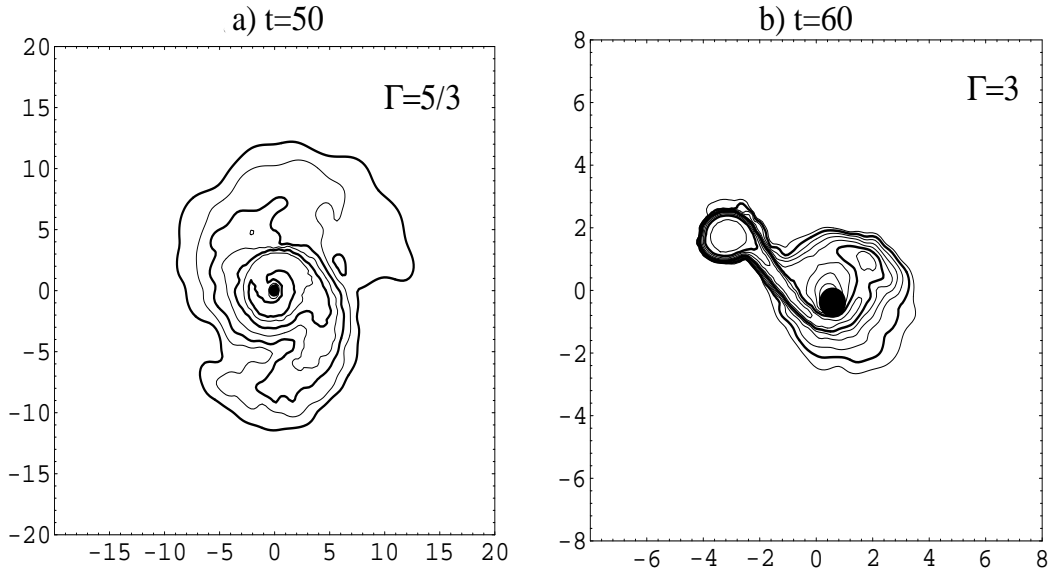,height=3.0in,width=5.5in}}
\vspace{6pt}
\caption{(a) Density contour plots in the orbital plane for run DL1.0
($\Gamma=5/3$) at $t=50$. All contours are logarithmic and equally
spaced every 0.5 dex. Bold contours are plotted at $\log
\rho=-5,-4,-3,-2$, in the units defined in section~\ref{initial}. (b) Same
as (a) but for run AL1.0 ($\Gamma=3$) at $t=60$, and with bold
contours plotted at $\log \rho=-3,-2,-1$.}
\label{fig5}
\end{figure}

By the same arguments as before, the outcome of these coalescence
events is reflected in the gravitational radiation signal as an abrupt
drop in amplitude (to practically zero) when the star is disrupted
(see Figure~4). Again the frequency at which this occurs will give a
measure of the radius of the neutron star.

\section{Discussion}

It is clear that there are serious limitations to the numerical
approach we have used to study this type of system, but we
nevertheless believe these simulations are useful for many
reasons. First, it is clear that hydrodynamical effects can play an
important role in the global behavior of the system at small
separations, and are crucial to determine the coalescence waveform
properly. Second, Newtonian calculations can be used to guide future,
more detailed simulations that will incorporate the effects of general
relativity, and can serve as useful guides in this respect.

The main result of all our calculations is that the outcome of the
coalescence process is very sensitive to the assumed stiffness of the
equation of state. For large values of the adiabatic index $\Gamma
\simeq 3$ the star is not disrupted, and there is a remnant left in
orbit around the black hole, with an accretion disk as well. The
gravitational radiation signal exhibits a drop in amplitude and a
return to lower frequencies, but does not vanish completely. The
coalescence process is delayed for at least several tens of
millisconds. For lower values of $\Gamma \simeq 2$, when the star does
not alter its radius in response to mass loss (or expands, when
$\Gamma<2$), complete tidal disruption is always observed, with
massive accretion disks forming around the black hole. The
gravitational radiation signal practically vanishes soon after Roche
lobe overflow occurs.

In a realistic scenario, the value of $\Gamma$ will not be uniform
throughout the star. However, for the purposes of tidal disruption and
the gravitational radiation emitted, it is the value at high densities
(roughly above nuclear density) that will determine the evolution of
the system. We have performed different tests using a variable
$\Gamma$ (it is specified as a function of density, examples of this
approach can be found in Rosswog et al.~2000), taking a stiff equation
of state for high densities, and a softer index for the low density
regions. The above discussion concerning tidal disruption is valid in
this case as well, if one considers the high--density value of the
adiabatic index, and since it is the bulk motion of matter that
determines the gravitational wave emission, the value of $\Gamma$ at
low densities is unimportant in this respect.

The approach one takes also clearly depends on the problem one wishes
to solve. These systems have been suggested as sources for the
production of cosmological gamma ray bursts (GRBs)
(Paczy\'{n}ski~1986; Goodman~1986; Eichler et al.~1989;
Paczy\'{n}ski~1991; Narayan, Paczy\'{n}ski \&
Piran~1992). Klu\'{z}niak \& Lee~(1998) showed that the conditions
during and after coalescence were indeed favorable for this, with the
creation of a thick accretion torus and a baryon--free axis in the
system, along the rotation axis, that would not hinder the production
of a relativistic fireball that could produce a GRB (M\'{e}sz\'{a}ros
\& Rees~1992, 1993). A different equation of state (that of Lattimer
\& Swesty~1991) has been used by Ruffert \& Janka~(1996) in the study
of double neutron star mergers, and by Janka et al.~(1999) in black
hole neutron star mergers. The stiffness of this equation of state is
not a free parameter, but there is a much greater level of detail in
the microphysics, relevant for the implications to GRB models (they
additionally included neutrino transport in their
calculations). Likewise, regarding the production of heavy elements
through r--process nucleosynthesis, a simple ideal gas treatment is
inadequate. One must use detailed thermodynamic calcualtions and a
realistic equation of state, as Rosswog et al.~(1999) and
Freiburghaus, Rosswog \& Thielemann~(1999) have done. Our computations
allow us simply to determine how much matter is dynamically ejected
from the system during coalescence, a necessary first step if it is to
contribute to the observed galactic abundances.

\section{Acknowledgements}

It is a pleasure to thank the organizers for a wonderful
workshop. This work was supported in part by CONACyT (27987E) and
DGAPA--UNAM (PAPIIT-IN-119998).

\end{document}